\documentclass[preprint,aps,prb,showpacs,amsmath,amssymb]{revtex4-1}
\usepackage{graphics}
\usepackage{epsf,graphicx,pstcol,epsfig,psfrag}
\begin{document}

\title{Isothermal magnetocaloric effect  in the vicinity of the Lifshitz point in Mn$_{0.9}$Co$_{0.1}$P}
\author{\frame{Tomasz Plackowski}, Marcin Matusiak, and Jozef Sznajd}
\affiliation{Institute of Low Temperature and Structural
Research, Polish Academy of Sciences, P.O. Box 1410, 50-950 Wroc³aw, Poland} 

\date{\today}

\begin{abstract}

The magnetic field - temperature ($B-T$) phase diagram of the Mn$_{0.9}$Co$_{0.1}$P single crystal is studied in the vicinity of the Lifshitz point by means of isothermal magnetocaloric coefficient ($M_T$) and AC susceptibility measurements. Results confirm previously reported shape of the $B-T$ phase diagram and locations of characteristic temperatures and fields.
At the Curie temperature ($T_C$) the critical exponent $\omega$, which describes a singularity of $M_T$ as a function of magnetic field ($M_T \propto B^{-\omega}$), is estimated for $B$ parallel to the easy axis to be equal to $\omega \approx 0.35$. Below $T_C$ an evidence of a new enigmatic phase, reported previously for pure MnP, is found in susceptibility data also for Mn$_{0.9}$Co$_{0.1}$P. However, the range of existence of this phase is significantly larger here, than in MnP. At the Lifshitz point we observe a sharp peak in the imaginary part of the magnetic susceptibility.
A phenomenological theory is introduced to describe the field dependence of the critical lines from the disordered phase (paramagnetic) to ordered phases (ferromagnetic and modulated). The temperature and field dependences of the magnetocaloric coefficient and susceptibility are also calculated within the same framework. 

\end{abstract}


\maketitle
\section{Introduction}
\label{introduction}
The magnetocaloric effect is the basis of the magnetic refrigeration technology, and manganese phosphide (MnP) was found to be a good candidate for such applications\cite{Reis}. However, the isothermal magnetocaloric effect can be also a useful tool, complementary to the specific heat and magnetic susceptibility measurements, for studying magnetic phase transitions \cite{Plac1, Plac2}. The technique for determining the isothermal magnetocaloric coefficient via precise measurements of the heat flux between the sample and its surrounding ($M_T \equiv \frac{\delta Q}{\delta B}$, where $Q$ is heat going out of the sample) was presented in Ref. \cite{Plac3}. In the present paper, we use this technique, along with measurements of the magnetic susceptibility, to study the high temperature part of the magnetic field - temperature $(B-T)$ phase diagram, thermodynamics and critical behavior of the Mn$_{0.9}$Co$_{0.1}$P single crystal. The phase diagram of Mn$_{0.9}$Co$_{0.1}$P which is a homologue of manganese phosphide with disorder on the metal sublattice was already described in several papers \cite{F84,F90,B90,Z92}. The high temperature part of this phase diagram was shown by means of AC susceptibility and magnetization measurements to be analogous to that of pure MnP. This includes a multicritical point which divides the critical line between disordered (Para) and ordered phases into two parts, such that on the first the phase transition to the ferromagnetic (Ferro), while on the second to modulated fan-like (Mod) is observed. There is a strong evidence that such a multicritical point in MnP exhibits Lifshitz type critical behavior \cite{Bec1,Bec2,Bec3,Bec4,Bec5}. 
The occurrence of a Lifshitz point in the $B-T$ phase diagram of MnP was discussed by Yokoi, Coutinho-Filho, and Salinas \cite{Yokoi}, who utilized a spin ($S=\frac{1}{2}$) localized model with competing interactions along the {\bf z} direction. On the basis of the mean-field approximation (MFA), with the assumption that the exchange parameters depend on $B$ and $T$, they calculated the transverse and longitudinal susceptibility of MnP asymptotically close to the Lifshitz point (LP). Zieba, Slota, and Kucharczyk \cite{Slota} considered axial next-nearest-neighbor Heisenberg (ANNNH) $S>\frac{1}{2}$ model and found, in the MFA, the ground state solutions. To extend the theory to finite temperature they assumed, similarly to the previous authors, that the model parameters are temperature dependent. For example, they chose arbitrary that the magnetization decreases with temperature according to the $\frac{3}{2}$ law, and in consequence the ratio $k=\frac{J_2}{J_1}$ of $J_1$ (connecting nearest neighbor layers) and $J_2$ (connecting next nearest neighbor layers) of the Heisenberg model changes with temperature in the following way:
\begin{equation}
k(T)=k(T=0)(1-c_k T^{\frac{3}{2}}),
\end{equation}
where $c_k$ was chosen to have $k(T_L=121K)=-\frac{1}{4}$.

Experimentally manganese phosphide has been studied from 1960's \cite{Huber} and its fundamental magnetic properties and richness of the magnetic phases, can be satisfactory explained in terms of competition between ferromagnetic and antiferromagnetic interactions \cite{Yos,Dob,Z92}. Despite the long history of MnP studies, in the year 2000 Beccera \cite{Bec2000} observed at very low field a novel phase below the ferromagnetic transition. Its existence has been recently confirmed only in one paper \cite{Yam}, but the nature of the phase remains unclear. This suggests that there is yet undiscovered physics, despite  MnP is considered as an archetypal magnetic system which exhibits the Lifshitz point. Similar behavior can be expected in Mn$_{0.9}$Co$_{0.1}$P, whose phase diagram closely resembles that of the pure MnP with the characteristic temperatures and fields scaled down \cite{B90,Z92}.
The global $B-T$ phase diagram of Mn$_{0.9}$Co$_{0.1}$P for $B \parallel$ {\bf b} was presented in Refs. \cite{B90} and \cite{Z92}. It was constructed on the ground of the susceptibility data. For 197.5 K $\ge$ T $\ge$ 100 K the authors found field dependences of the susceptibility characteristic for the Para/Ferro phase transition, whereas for $ T < 100 K$ the transition to the Para phase occurred in two steps. The phase between the Para and Ferro phases
was identified as the fan phase by analogy with MnP. The critical fields were determined by the intersection of tangents to the $\chi'(B)$ curves, which were, as the authors stated, to some extent arbitrarily selected just below and above the transition \cite{Z92}.

The main purpose of this paper is to report the results of the magnetocaloric experiments for Mn$_{0.9}$Co$_{0.1}$P and show a usefulness of this technique to study phase diagrams and critical behavior of magnetic systems in the field. We also investigate the possibility of the occurrence of the novel phase, which was reported by Beccera \cite{Bec2000} in MnP, in the Co-diluted system. Finally, we propose a simple phenomenological "hybrid" theory which allows us to reconstruct the high temperature part of the $B-T$ phase diagram as well as temperature and field dependences of magnetocaloric coefficient, magnetization and transverse susceptibility in the vicinity of order - disorder phase transitions lines in magnetic systems which exhibit the Lifshitz multicritical point.

\section{Experiment}
\label{Experiment}
We study the Mn$_{0.9}$Co$_{0.1}$P single crystal that was obtained by the Bridgman method as described in Ref.\cite{F90}. The same sample was already used in previous investigations by A.~Zieba et al. in Ref.\cite{Z92}. X-ray-diffraction studies showed no signs of any long- or short-range ordering of the Co and Mn atoms, what was confirmed by neutron-diffraction studies of powder samples. The possibility of macroscopic concentration gradients was excluded by performing x-ray-fluorescence scans on the different faces of the cube sample with no indication for change in the [Mn]:[Co] ratio \cite{Z92}.

The isothermal magnetocaloric coefficient and specific heat data were obtained using a heat-flow calorimeter  \cite{Plac3}. In this method a sample is connected to the heat sink by means of the sensitive heat-flow meter of high thermal conductance. The sample was glued to the heat-flow meter using Collaprene (Gubra, Milan, Italy) with the either {\bf b} (intermediate) or {\bf c} (easy) axis oriented parallel to the magnetic field. The sample was surrounded by a double passive radiation screen (gold plated), and both screens were in a good thermal contact with the sink. The whole ensemble was evacuated down to $10^{-4}$ Pa and placed in the gas-flow variable-temperature insert of the Oxford Instruments cryostat fitted with a 13/15 T superconducting magnet. Temperature dependences of the AC magnetic susceptibility in a magnetic field were measured using a Quantum Design Physical Property Measurement System (PPMS).

\section{Results}
\label{Results}
The motivation for choosing the Mn$_{0.9}$Co$_{0.1}$P single crystal was influence of the cobalt substitution on the Curie temperature ($T_C$) in this compound. Namely, 10\% Co-doping lowers the Curie temperature from $T_C$ = 291.5 K in MnP \cite{Huber} down to 199 K \cite{Z92} in Mn$_{0.9}$Co$_{0.1}$P. $T_C$ in undoped MnP lies just on the border of our experimentally accessible temperature region, therefore we decided to investigate Mn$_{0.9}$Co$_{0.1}$P to avoid possible difficulties. On the other hand, one can expect that Mn$_{0.9}$Co$_{0.1}$P remains a good qualitative analog of the undoped manganese phosphide, since cobalt acts as a non-magnetic diluent, where Mn and Co atoms are randomly distributed with no superstructures \cite{Z92}. The magnetic susceptibility data for $B \parallel {\bf b}$ presented in figures 1 and 2 confirm these expectations.
\begin{figure}
\label{Fig1}
 \epsfxsize=10cm \epsfbox{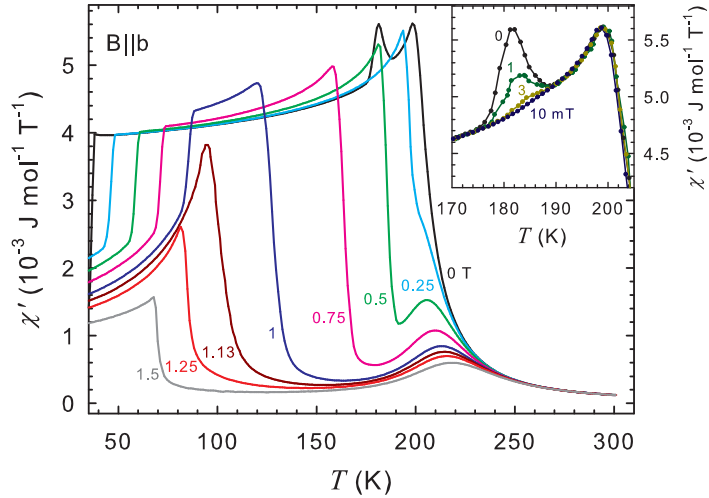}
 \caption{(Color online) Temperature dependences of the real part of the AC susceptibility $\chi'(T)$ for various magnetic fields up to $B$ = 1.5 T applied along the {\bf b} axis. The modulation field was $B_{AC}\parallel {\bf b}$ = 1 mT, with frequency $f_{AC}$ = 1011 Hz. Inset shows vicinity of the Para/Ferro transition with an additional peak at $T\approx 180$ K.}
 \end{figure}
For example an enigmatic magnetic phase, which was reported to occur in MnP below $T_C$ at very low magnetic field \cite{Bec2000,Yam}, seems to be also present in Mn$_{0.9}$Co$_{0.1}$P. The corresponding phase transition manifests itself as a peak below the Ferro/Para transition in both real ($\chi'(T)$ - see inset in Fig. 1) and imaginary ($\chi''(T)$ - see Fig. 2) part of the AC magnetic susceptibility. In zero field the peak in $\chi''(T)$ is about two orders of magnitude higher than the anomaly at $T_C$ .   
\begin{figure}
\label{Fig2}
 \epsfxsize=10cm \epsfbox{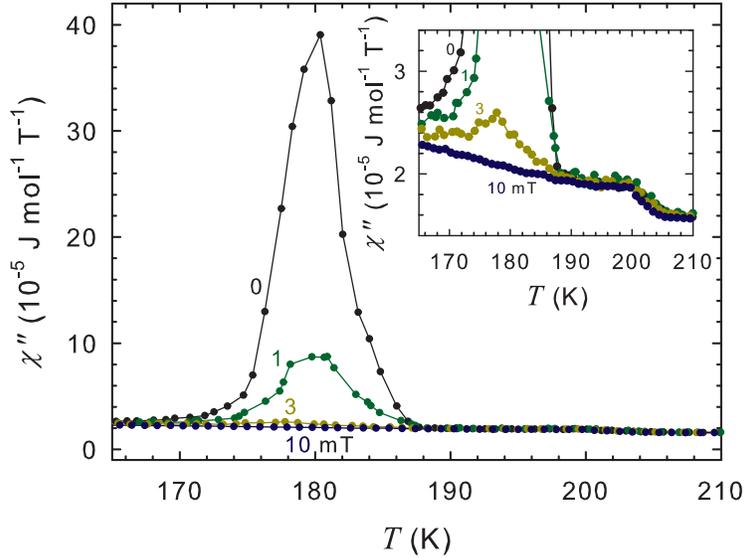}
 \caption{(Color online) Temperature dependences of the imaginary part of the AC susceptibility $\chi''(T)$ for various magnetic fields ($B$ = 0,1,3,10 mT) applied along the ${\bf b}$ axis. The modulation field was $B_{AC} \parallel {\bf b}$ = 1 mT, with frequency $f_{AC}$ = 1011 Hz.}
 \end{figure}
The magnetic field of approximately 5 mT suppresses anomalies in both $\chi'$ and $\chi''$. This means that the low-field phase in Mn$_{0.9}$Co$_{0.1}$P is significantly more field-resistant than in undoped MnP, where the suppressing field was one order of magnitude smaller\cite{Bec2000,Yam}. Hence cobalt doped Mn$_{1-x}$Co$_{x}$P could be a good candidate to investigate properties of the new magnetic phase.
Another characteristic that is common for Mn$_{0.9}$Co$_{0.1}$P and MnP is a broad maximum in $\chi'(T)$ dependences that appears above Para/Ferro transition for non-zero magnetic field (see Fig. 1). Its emergence was suggested to be a result of critical fluctuations preceding the Para/Ferro transition, but we believe it can be of different origin as discussed later in "The Model" section.

A feature that has not been reported previously for MnP is a peak in the imaginary part of the magnetic susceptibility at the Lifshitz Point. The coordinates of LP are $B\approx$ 1.1 T, $T\approx$ 95 K, and the peak in $\chi''(T)$ vanishes for higher and lower temperatures (see main panel of Fig. 3), as well as higher and lower magnetic fields (see inset in Fig. 3).
\begin{figure}
\label{Fig3}
 \epsfxsize=10cm \epsfbox{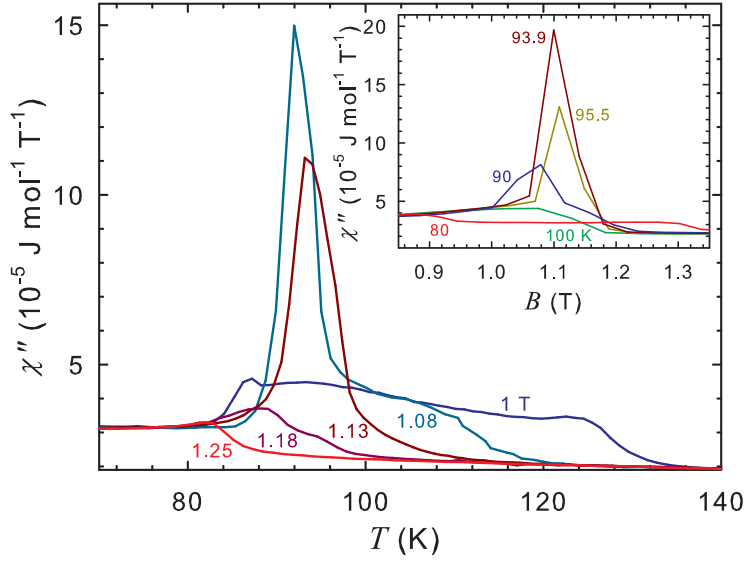}
 \caption{(Color online) Temperature dependences of the imaginary part of the AC susceptibility $\chi''(T)$ in the vicinity of the Lifshitz Point for various magnetic fields ($B \parallel {\bf b}$). Inset shows  $\chi''(B)$ field sweeps for several temperatures around LP.}
 \end{figure}
The increase of AC losses was already observed at the Mod/Ferro transition line and recognized as a sign of a discontinues phase transition \cite{Bec6}. However, a size of the peak at the Mod/Ferro line is about two orders of magnitude smaller than at LP. Additionally, a position of the peak in  $\chi "(T)$ is almost independent of the AC field frequency - the temperature shift is only about 1 K between 50 and 10 000 Hz. We are not certain whether this significant rise of the AC losses at the Lifshitz Point is generally related to its unique critical properties, or this occurs specifically in Mn$_{0.9}$Co$_{0.1}$P.

The present paper is focused mainly on the thermodynamic properties of Mn$_{0.9}$Co$_{0.1}$P and our primary tool are measurements of 
the isothermal magnetocaloric coefficient that can be defined as $M_T=-T (\delta S/ \delta B)\mid_T$, where $S$ is the entropy of the system.
 Several representative $M_T$($B$) curves for $B \parallel {\bf b}$ are presented in figure 4.
\begin{figure}
\label{Fig4}
 \epsfxsize=10cm \epsfbox{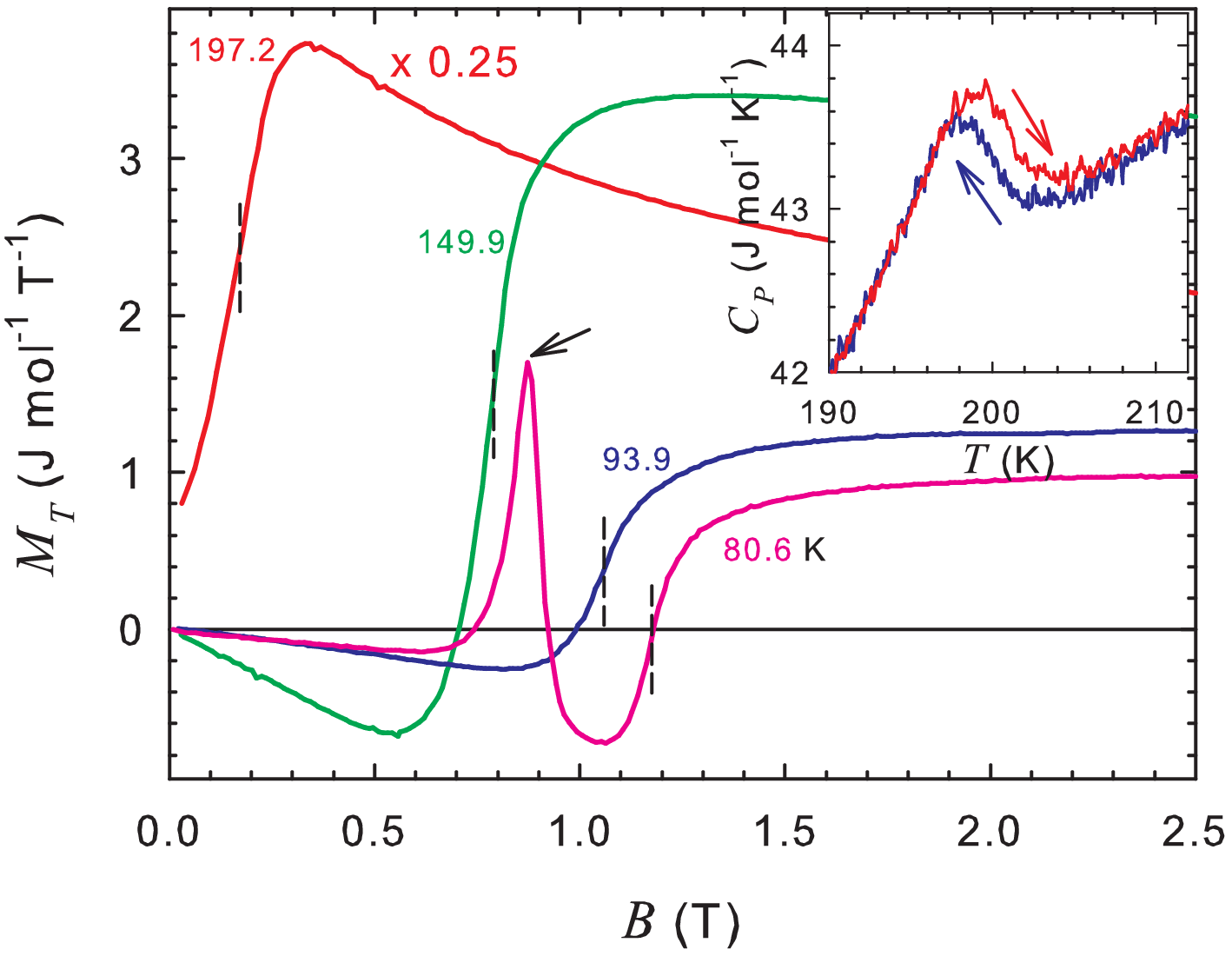}
 \caption{(Color online) Field dependences of the isothermal magnetocaloric coefficient for various temperatures: $T=197.2$ K (at $T_C$), $T=149.9$ K (at the step-like Para/Ferro transition), $T=93.9$ K (at LP), and $T=80.6$ K (where two transitions are visible: 1st order Ferro/Mod indicated by arrow, and step-like 2nd order Para/Mod). Dashed lines indicate the $M_T$($B$) inflection points. All curves are measured in decreasing magnetic field $B \parallel {\bf b}$. Inset shows temperature dependence of the specific heat ($B=0$ T) in the vicinity of $T_C$.}
 \end{figure}
The isothermal magnetocaloric effect measurements allow to investigate properties of horizontal-like transition lines on the $B-T$ phase diagram, whereas the specific heat measurements are useful in studying vertical-like lines. Therefore, both methods are complementary. In Mn$_{0.9}$Co$_{0.1}$P a predominant part of Para/Ferro transition line is level (see Fig. 5), and as a consequence the anomaly in the specific heat related to Para/Ferro transition measured at constant magnetic field becomes practically undetectable for B $\geq$ 0.5 T. In contrast, a step-like anomaly in $M_T(B)$ is clearly visible in the entire region of occurrence of the Para/Ferro transition.
A difficulty is related to the fact that there is no characteristic point on $M_T$($B$) that can be unambiguously recognized as the Para/Ferro transition field ($B_C$) \cite{Sznajd}. In our opinion a good candidate to define $B_C$ is an inflection point that is present in all $M_T$($B$) curves as shown in figure 4. The transition lines determined in this way coincide satisfactorily with previously reported results of magnetic measurements \cite{Z92} denoted by red dashed line in figure 5. 
\begin{figure}
\label{Fig5}
 \epsfxsize=10cm \epsfbox{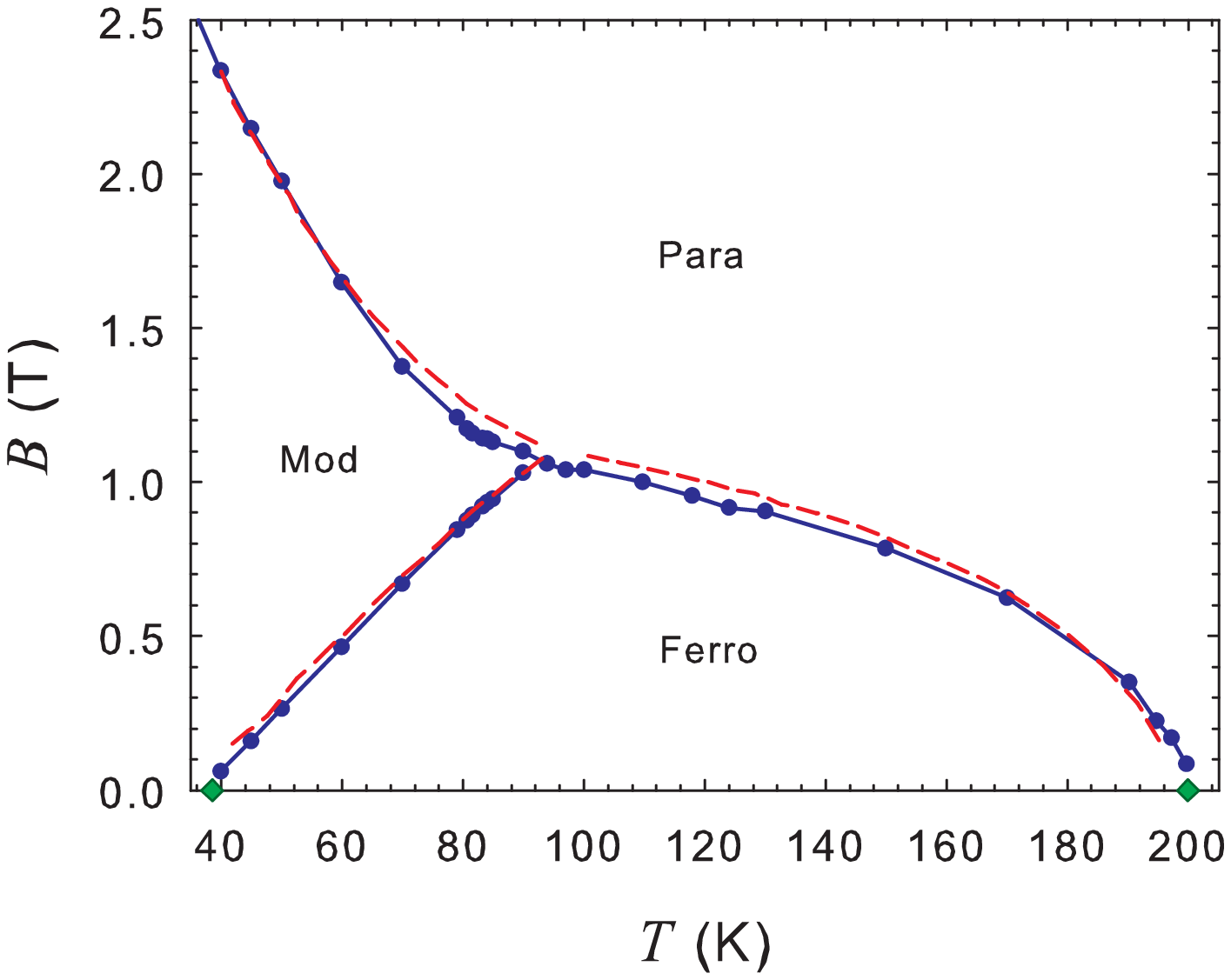}
 \caption{(Color online) $B-T$ phase diagram of Mn$_{0.9}$Co$_{0.1}$P ($B \parallel {\bf b}$) determined on the basis of $M_T$ (blue points) and $C_P$ (green diamonds) measurements. The red dashed line depicts results of magnetic susceptibility studies reported in Ref. \cite{Z92}.}
 \end{figure}
For $B \parallel {\bf c}$ (easy axis) $M_T$($B$) diverges at the Curie temperature as a power law in $B$ $M_T(B) \propto B^{-\omega}$ ($\omega$ is the critical exponent) as shown in figure 6. For temperature higher and lower than $T_C$ and also for $B \parallel {\bf b}$ this trend breaks when nearing $B=0$. A small deviation of $M_T$($B$) from the power dependence, seen at low field also for $T=199.9$ K and $B \parallel {\bf c}$, is probably caused by a tiny difference between $T_C$ and the actual temperature of the measurement.
\begin{figure}
\label{Fig6}
 \epsfxsize=10cm \epsfbox{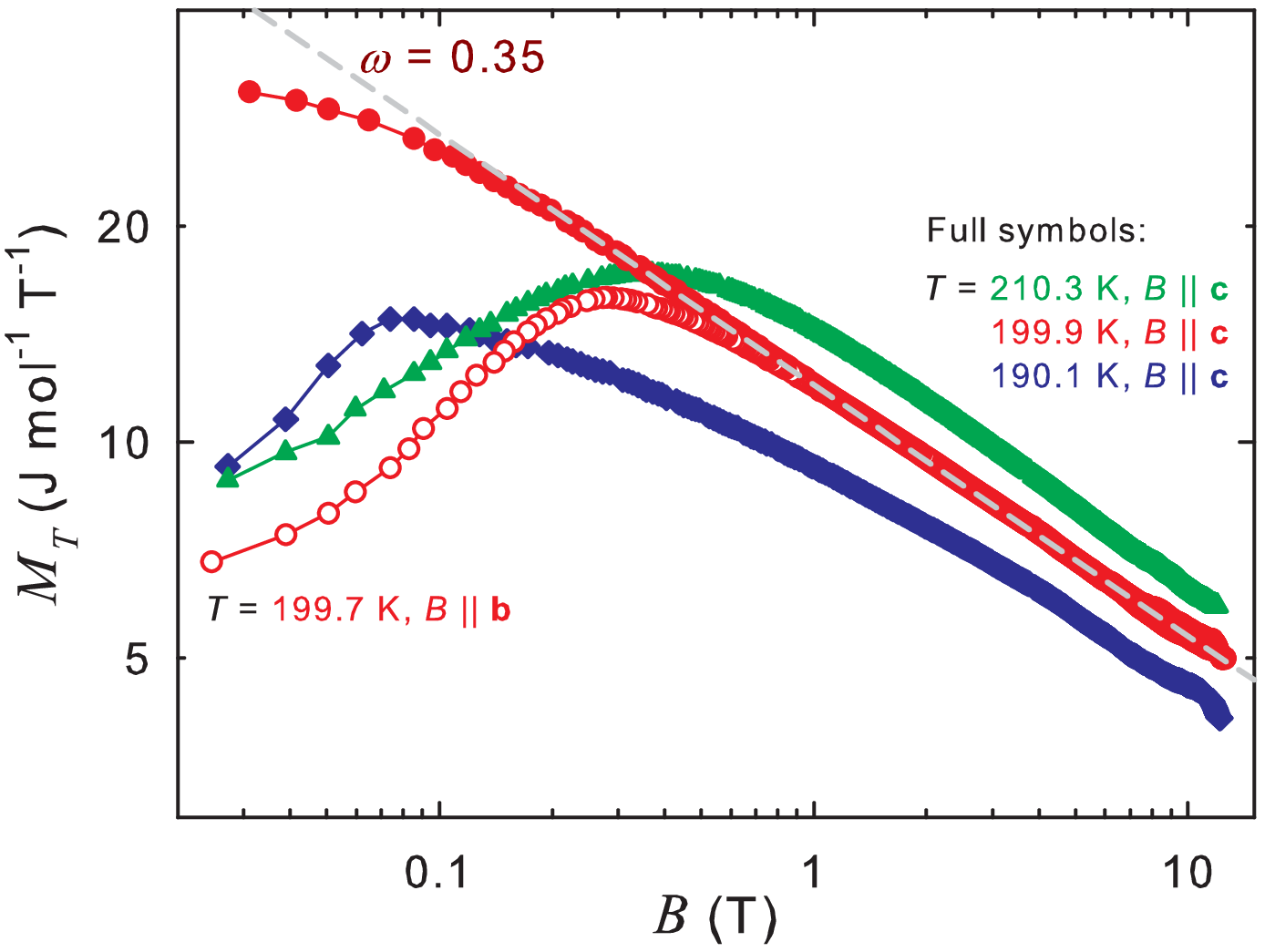}
 \caption{(Color online) Isothermal magnetocaloric coefficient in the vicinity of the Curie temperature ($T_C \approx 200$ K) plotted against magnetic field in logarithmic scale. Full symbols denote data for $B \parallel {\bf c}$  , whereas open points are for $B \parallel {\bf b}$ configuration. Dashed line denotes a $M_T(B) \propto B^{-\omega}$ fit.}
 \end{figure}
The $\omega$ exponent obtained from fitting $M_T(B)$ at $T_C$ equals 0.35. This can be compared with results of magnetization studies of MnP \cite{Terui} by using equation $ \omega = \frac{1-\beta}{\beta  \delta}$ \cite{Plac4}, where $\beta$ is the exponent of temperature-, whereas $\delta$ field-, -dependent magnetization. Namely, for MnP and the easy axis Terui et al. obtained $\beta=0.34 \pm 0.02$ and $\delta=4.89 \pm 0.1$, what gives $\omega \approx 0.4$. This value would indicate that critical behavior of the system is located between three dimensional Ising ($\omega=0.433$) and three dimensional Heisenberg ($\omega=0.364$) universality class. Our value of $\omega=0.35$ at $T_C$ would suggest that in case of cobalt doped Mn$_{0.9}$Co$_{0.1}$P the isotropic 3D Heisenberg model is more appropriate. 
The critical exponents at lower temperatures, where the transition occurs in non-zero magnetic field, can be, in principle, retrieved using the Riedel and Wegner prediction that the transverse susceptibility at constant temperature diverges at the transition as $\chi \sim \mid B-B_C \mid^{-\alpha}$ \cite{Riedel} ($\alpha$ is specific heat critical exponent). Bindilatti et al. \cite{Bec4}, who tried to determine $\alpha$ from the study of the magnetic susceptibility, found $\alpha^- = 0.39$ significantly different from $\alpha^+ = 0.53$ and much higher than theoretical estimate ($0.18-0.25$) \cite{Pleimling,Diehl,Leite}. The authors claimed that the observed discrepancies in $\alpha$ values are caused by the closeness of the discontinuous Ferro/Mod transition. According to the Riedel and Wegner \cite{Riedel} the same type of the singularity should be also observed for $M_T$ as a function of $(B-B_C)$. However, a residue of a singular part can be small and difficult to observe. In fact, it is not possible to determine the power law behavior of $M_T$($B$) singular part for $B \parallel ${\bf b} close to LP, because the anomaly in $M_T$($B$) connected with the Para/Ferro transition has the step-like shape (see Fig. 4). On the other hand, it should be also noticed that one of the Riedel and Wegner \cite{Riedel} assumptions: $T_C(B)-T_C(B=0$ T$) \sim B^2$, which is true for the small field, is not fulfilled in the vicinity of the LP (see Fig. 5). It leads us to the conclusion that, in general, one cannot find the specific heat critical exponent $\alpha$ from the transverse susceptibility or magnetocaloric measurements. It is despite the fact that within the Landau theory all three quantities: specific heat, transverse susceptibility and magnetocaloric coefficient, share the same behavior, i.e. have a jump at the transition line:

\begin{equation}
C^P-C^F =\frac{1}{t_c} \frac{a^2 k^4}{b^2 h^2} ( \chi^P-\chi^F)  = \frac{1}{t_c} \frac{a k^2}{b h}  (M_T^P-M_T^F). 
\end{equation}
This is based on the Landau free energy expressed in the form:
\begin{equation}
f_L=a (t - t_c) m_z^2 + (a (t - t_c) - k) m_x^2 + b (m_x^2 + m_z^2)^2 + h m_x,
\end{equation}
where $t_c$ is the zero-field critical temperature, $k$ uniaxial anisotropy constant, and $m_z, m_x$ magnetization components along and perpendicular to the easy axis, respectively.

\section{The model}
A complete microscopic theory for magnetic behavior of MnP and all the more of Mn$_{0.9}$Co$_{0.1}$P is not available. The most often used and the most fruitful one is based on the localized spin model with competing effective nearest-neighbor and next-nearest-neighbor interactions \cite{Hiyamizu,Yokoi,Slota}. The zero temperature solution of ANNNH model \cite{Slota} explains some experimental results observed in MnP. However, to describe the behavior of the system in the vicinity but a finite distance away from LP, the authors had to choose arbitrary the dependence of the model parameter on temperature \cite{Slota}, or on temperature and field \cite{Yokoi}. Because of the quenched disorder the detailed theoretical interpretation of the phenomena observed in Mn$_{0.9}$Co$_{0.1}$P based on the microscopic model is, of course, more complicated. 
Therefore, in order to describe the main thermodynamic features of Mn$_{0.9}$Co$_{0.1}$P in the vicinity of the Lifshitz point, we consider a very simple phenomenological model based on the localized spin-$\frac{1}{2}$ model Hamiltonian. The model considers spin layers with anisotropic ferromagnetic intralayer interactions described by the Hamiltonian
\begin{eqnarray}
\label{1} 
{\cal H} &=& - \sum_{\alpha=x,y,z}\sum_{i,j,n} j_{\alpha} S_{\alpha,i}^{n} S_{\alpha,j}^{n}  - h \sum_{i,n} S_{z,i}^n,
\end{eqnarray}
where $ S_{\alpha,i}^{n}$ denotes the $i-th$ $S=\frac{1}{2}$ spin in $n-th$ layer,  $h\equiv B/\mu$. and $j_x>j_y>j_z$. The easy axis is along $x$ direction, the $z$ and $y$ axes are the medium- and hard magnetization direction respectively. The magnetic moment near the critical line lies in the $(x,z)$ plane. The ferromagnetic layers are coupled by interlayer interactions
\begin{eqnarray}
\label{2} 
{\cal H}_{I} &=& - \sum_{i,n,p} j_{x}^{(p)} S_{x,i}^{n} S_{x,i}^{n+p}+{\cal H}_q.
\end{eqnarray}
The phase diagram of Mn$_{0.9}$Co$_{0.1}$ and magnetic structure of the low temperature phases was presented in previous papers \cite{F90,B90,Z92}. Here, our aim is to describe the thermodynamic behavior near LP, so we take into account only the interaction between $x$ spin-components connecting the $n-th$ layer to $(n+p)-th$ layers and we do not assume the explicit form of the ${\cal H}_q$ interaction. For the thermodynamics of the Para/Ferro phase transition the second term in (5) plays no role and by using (4) and (5) one can easily find, in the MFA, the Landau free energy in the following form
\begin{eqnarray}
\label{3} 
f_L &=&f_0+a_x m_x^2+b_x m_x^4+c_z m_z+a_z m_z^2+r_z m_z^3+b_z m_z^4+d m_x^2 m_z+d_x m_x^2 m_z^2,
\end{eqnarray}
where:
\begin{eqnarray}
\label{4}
f_0 &=& -t \log {2 \cosh{\frac{h}{t}}}, \quad a_x=J_x(1-\frac{2 J_x \tanh{\frac{h}{t}}}{h}),\quad 
b_x=\frac{J_x^4 sech{(\frac{h}{t})^2}(t \sinh{\frac{2 h}{t}})}{h^3 t} \nonumber \\
c_z &=& -2 J_z \tanh{\frac{h}{t}}, \quad a_z=\frac{J_z(2 J_z \tanh{(\frac{h}{t})^2-2 J_z+t)}}{t}, \quad r_z=\frac{8 J_z^3 sech{(\frac{h}{t})^2}  \tanh{\frac{h}{t}}} {3 t^2} \nonumber \\
b_z&=&\frac{4 J_z^4(2-\cosh{\frac{2 h}{t}}) sech{(\frac{h}{t})^4}}{3 t^3}, \quad d=\frac{4 J_x^2 J_z(t e^{\frac{4 h}{t}}-4 h e^{\frac{2 h}{t}}-t)}{(1+e^{\frac{2 h}{t}})^2 h^2 t}, \nonumber \\ d_x &=& \frac{8 J_x^2 J_z^2(t^2 e^{\frac{6 h}{t}}-t^2+e^{\frac{2 h}{t}}(4 h^2-4 h t-t^2)+e^{\frac{4 h}{t}} (t^2-4 h^2-4 h t))}{(1+e^{\frac{2 h}{t}})^3 h^3 t^2}.
\end{eqnarray}
Taking into account the exchange constants connecting only nearest- and next-nearest-neighbor layers we have: 
\begin{eqnarray}
\label{5} 
J_x=z_0 j_x+z_1 j_x^{(1)}+z_2 j_x^{(2)}, \quad J_z=z_0 j_z,
\end{eqnarray}
where $z_{i}$ denote the appropriate coordination  numbers. Thus, in the present approximation, the system is in fact described by only one internal parameter $J_z$ (one can assume $J_x=1$), reduced temperature $t$, and external field $h$, which are measured in units of $J_x$. 

By minimizing the free energy (6) one can easily find the paramagnetic phase with magnetization along the field $m_x=0$, and $m_z$ being the solution of the following cubic equation
\begin{eqnarray}
\label{6} 
c_z+2 a_z m_z+3 r_z m_z^2 +4 b_z m_z^3=0,
\end{eqnarray}
which can be easily found in the form:
\begin{eqnarray}
\label{7} 
m_z=-\frac{r_z}{4 b_z}-\frac{24 a_z b_z-9 r_z^2} {6 \sqrt[3]{4} b_z W}+\frac{W}{12 \sqrt[3]{2} b_z},
\end{eqnarray}
where
\begin{eqnarray}
\label{8} 
W=\sqrt[3]{(2 \sqrt{(24 a_z b_z-9 r_z^2)^3+729 (8 b_z^2 c_z-4 a_z b_z r_z+r_z^3)^2}-432 b_z^2 c_z+216 a_z b_z r_z-54 r_z^3)}.
\end{eqnarray}
In the ferromagnetic phase ($m_x \neq 0$ and $m_z \neq 0$) $m_z$ is also a solution of the cubic equation with slightly more complicated coefficients:
\begin{eqnarray}
\label{9} 
c_z-\frac{a_x d}{2 b_x}+(2 a_z-\frac{d^2+2 a_x d_x}{2 b_x}) m_z+3 (r_z-\frac{d d_x}{2 b_x}) m_z^2+(4 b-\frac{d_x^2}{b_x}) m_z^3=0,
\end{eqnarray}
and
\begin{eqnarray}
\label{10} 
m_x=\sqrt{-\frac{a_x+d m_z-d_x m_z^2}{2 b_x}}.
\end{eqnarray}
The temperature dependences of the Para/Ferro phase transition order parameters $m_x$ (magnetization component perpendicular to the field direction) and $m_z$ (magnetization component along the field) for $J_z=0.8$ and $h=0.15$ are presented in Fig.7.
\begin{figure}
\label{Fig7}
 \epsfxsize=10cm \epsfbox{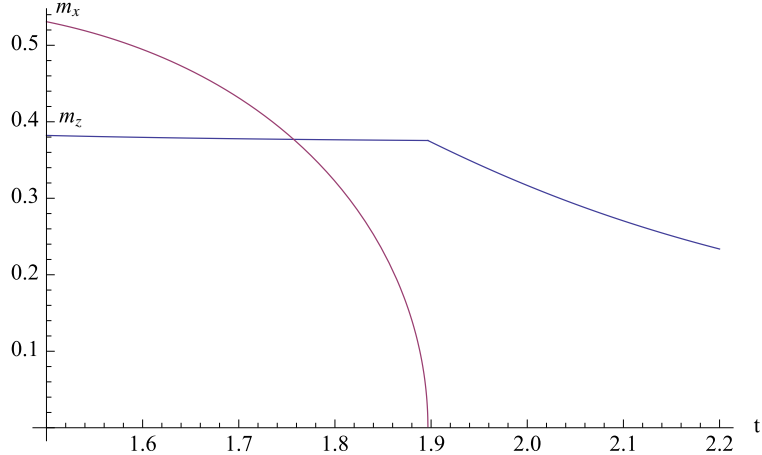}
 \caption{(Color online) Temperature dependences of the Para/Ferro phase transition order parameter - $m_x$ (red line) and magnetization component along the field $m_z$ (blue line) for $J_z=0.8$ and $h=0.15$ .}
 \end{figure}

Fig.8 shows the temperature dependence of the magnetic susceptibility $\chi=\frac{\delta m_z}{\delta{h}}$ in the vicinity of the Para/Ferro phase transition for $J_z=0.8$ and several values of the external magnetic field.

\begin{figure}
\label{Fig8}
 \epsfxsize=10cm \epsfbox{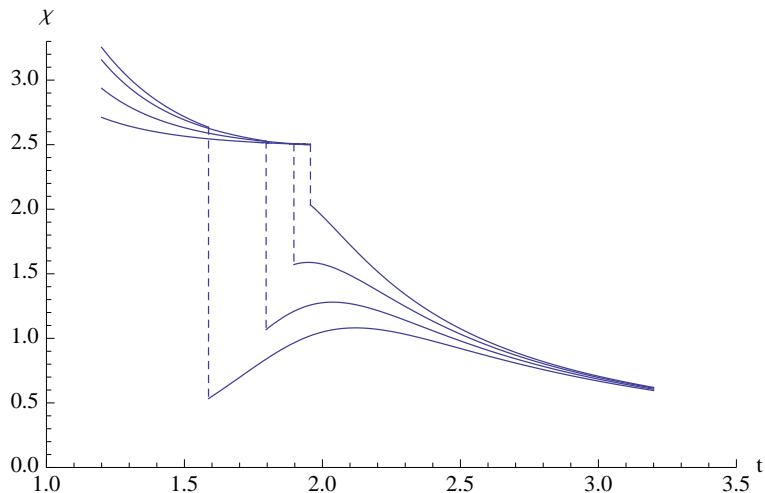}
 \caption{Temperature dependences of the susceptibility for $J_z=0.8$ and $h=0.1, 0.15, 0.2$, and $0.25$ (Para/Ferro phase transition) from top to the bottom.}
 \end{figure}

As seen in the paramagnetic phase $\chi$ has a maximum that decreases, widens and shifts to higher temperature with increasing field. Such a maximum is visible in our measurements (Fig. 1) and was previously reported by Beccera \cite{Bec2000}, who claimed that the existence of this maximum "is explained in terms of critical fluctuations that precede the ferromagnetic transition". However, in the present approximation (MFA) critical fluctuations are neglected and we obtain the similar behavior of $\chi$ nevertheless. It suggests that perhaps the explanation of the susceptibility temperature dependence of the system under consideration does not require to take into account the order parameter fluctuations. It is worth emphasizing that usually the existence of the susceptibility maximum in the paramagnetic phase of ferromagnets in a field is not connected with the critical fluctuations and can be observed even in the systems which do not undergo any phase transition \cite{Sz}. Such a maximum is characteristic (but not universal in the sense of the power law dependence of the maximum temperature location $t_m$ on a field \cite{Sz,Chen}) feature of the magnetic systems with ferromagnetic interactions in the phase in which the magnetization is parallel to the field.

In order to analyze the Para/Mod phase transition near the multicritical point we insert two following terms into free energy (6) in the spirit of the Landau theory
\begin{eqnarray}
\label{11} 
A_q m_x^2 q^2+B_q m_x^2 q^4,
\end{eqnarray}
where $q$ is additional, fictitious order parameter that describes the difference between the magnetization of the two adjacent layers. The parameter $q$ is different from zero only if the order parameter $m_x\neq0$. Now, by minimizing the free energy (6) supplemented by the two terms  of (14)
\begin{eqnarray}
\label{12} 
f_{L_q}=f_L+A_q m_x^2 q^2+B_q m_x^2 q^4,
\end{eqnarray}
and except for the uniform solution $q=0$, one can find the solution describing a modulated phase with
\begin{eqnarray}
\label{13} 
q=\sqrt{-\frac{A_q}{2B_q}},
\end{eqnarray}
and magnetic order parameter
\begin{eqnarray}
\label{14} 
m_x=\sqrt{\frac{A_q^2-4 a_x B_q-4 B_q d m_z-4 B_q d_x m_z^2}{8B_q b_x}},
\end{eqnarray}
where similarly to the previous cases $m_z$ is a solution of the cubic equation
\begin{eqnarray}
\label{15} 
c-\frac{a_x d}{2 b_x}+\frac{a^2 d}{8 B_q b_x}+(2 a_z-\frac{d^2}{2 b_x}-\frac{a_x d_x}{b_x}+\frac{A_q^2d_x}{4 B_q b_x}) m_z+3 (r_z-\frac{d d_x}{2 b_x}) m_z^2+(4 b-\frac{d_x^2}{b_x}) m_z^3=0.
\end{eqnarray}
It is easy to see that for $A_q<0$ the ferromagnetic phase with $m_x\neq0$ and $q=0$ is unstable with regards to $q$ because
\begin{eqnarray}
\label{13} 
\frac{\delta^2 f_{L_q}}{\delta q^2}=2m_x^2(A_q+6 B_q q^2) < 0.
\end{eqnarray}
Unfortunately, in contradistinction to the coefficients of the free energy (6) we do not know the form of the coefficients $A_q$ and $B_q$ as functions of the microscopic parameters. Thus according to the Landau theory, we assume that $B_q$ is constant near the LP and $A_q$ is given by the formula 
\begin{eqnarray}
\label{13} 
A_q=A'(t-t_{LP}),
\end{eqnarray}
where $t_{LP}$, the critical temperature of the LP transition, depends on $J_z$ and $h$ and is the end point of the Para/Ferro critical line. For $J_z=0.8$ and $h=0.25$, $t_{LP} \approx 1.59$.  
\begin{figure}
\label{Fig9}
 \epsfxsize=10cm \epsfbox{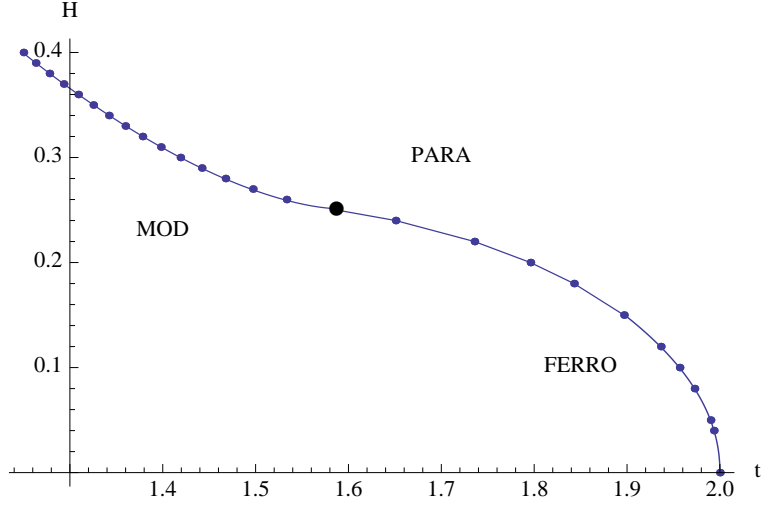}
 \caption{Phase diagram (H,t) for model (1) with the critical order/disorder line divided by the multicritical point (full circle). Critical line Para/Ferro for $h<0.15$, $h_c=0.44(2-t)^{0.475}$, Para/Mod $h_c=0.25+0.84 (1.59-t)^{1.6}$.}
 \end{figure}
In Fig.9 we present the high temperature part of the phase diagram, namely the critical lines between the Para and ordered phases for $J_z=0.8$ and $\frac{A'}{B_q}=2$ with the multicritical point at ($h=h_{LP}=0.25, t=t_{LP}\approx1.59$). For a  field small enough, i.e. $h<0.15$, the critical line Para/Ferro can be satisfactory fitted to the formula
\begin{eqnarray}
\label{13} 
h_c=0.44 (2 - t)^{\lambda}, \quad \lambda \approx 0.475
\end{eqnarray}
 with $\lambda$ close to the asymptotic value for $h->0$, $\lambda=0.5$. It is worth nothing that for the transition Para/Mod the convexity of the critical line changes (according to the experimental results) and it can be fitted to 
\begin{eqnarray}
\label{13} 
h_c=0.25+0.84 (1.59 - t)^{\lambda}, \quad \lambda \approx 1.6.
\end{eqnarray}
In Fig.10 the temperature dependences of the order parameter $m_x$, fictitious order parameter $q$, and magnetization component along the field $m_z$  for $J_z=0.8$ and $\frac{A'}{B_q}=2$ at the external magnetic field $h=h_{LP}$ are presented. As observed experimentally in the Mod phase $m_z$ remarkable decreases with decreasing temperature whereas in the Ferro phase (see Fig.7) is almost temperature independent.
\begin{figure}
\label{Fig10}
 \epsfxsize=10cm \epsfbox{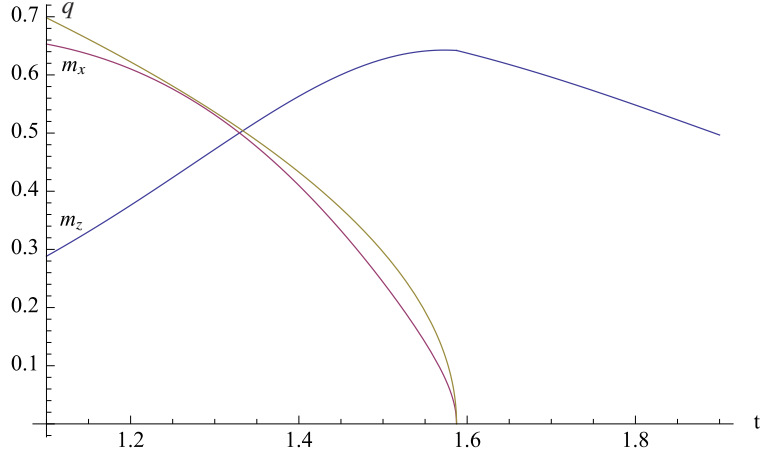}
 \caption{(Color online) Temperature dependence of the order parameters $m_x$ (red line), $q$ (yellow line) and $m_z$ (blue line) at the Lifshitz point field $h=h_{LP}$.}
 \end{figure}

Fig.11 shows the temperature dependence of the longitudinal susceptibility for a field $h\ge h_{LP}$ (Para/Mod phase transition). As seen in Fig.10 similarly as for the fields $h\le h_{LP}$ (Fig.8) also here the maximum of the susceptibility exists and its location is shifted towards higher temperature with increasing field. However, the temperature dependence of $\chi$ below the phase transition is different in both cases (compare the Figs. 8 and 10, and experimental results, Fig. 1).  
\begin{figure}
\label{Fig11}
 \epsfxsize=10cm \epsfbox{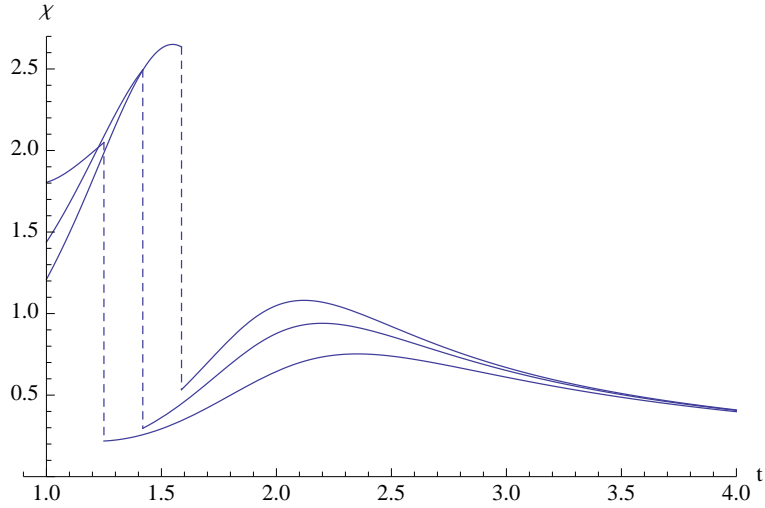}
 \caption{Temperature dependences of the longitudinal susceptibility for $J_z=0.8$ and $h=0.25, 0.3$, and $0.4$ (Para/Mod phase transition) from top to the bottom.}
 \end{figure}

\begin{figure}
\label{Fig12}
 \epsfxsize=10cm \epsfbox{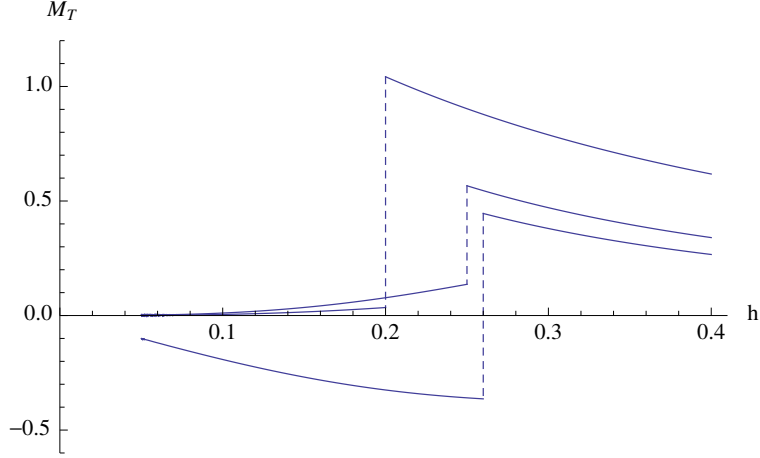}
 \caption{Field dependence of the isothermal magnetocaloric coefficient for $J_z=0.8$ and $t =$ 1.5333, 1.5874 (LP), 1.796 from bottom to the top.}
 \end{figure}

In Fig.12 the isothermal magnetocaloric coefficient $M_T$ as a function of field is given for temperature above LP ($t\approx1.53$) - top curve, close to the LP ($t\approx1.59$) - intermediate curve, and above LP ($t\approx 1.8$) - bottom curve. As seen for temperatures below the Lifshitz Point, $M_T$ changes sign at the critical field. This phenomenon is observed also in the experiment (Fig. 3) and in consequence the critical lines of the Para/Mod transition found from the sign changes of $M_T$ and from an inflection point  of the curve $M_T(h)$ coincide. It should be emphasized that this is not the case for the Para/Ferro transition. As seen in Fig.3 $M_T$ is negative also in the ferromagnetic phase near the LP, which testifies to the existence of the antiferromagnetic correlation in this region \cite{Sznajd}.

\section{Conclusion}

Manganese phosphide is a paradigm of a magnetic system in which near the confluence of the ordered ferromagnetic and modulated phases with the paramagnetic phase, the LP critical behavior can be experimentally explored. Similar physics is observed in Co-diluted Mn$_{0.9}$Co$_{0.1}$P with the characteristic temperatures and fields scaled down. Measurements of the isothermal magnetocaloric coefficient presented in this paper have confirmed the shape of the Mn$_{0.9}$Co$_{0.1}$P phase diagram and the location of the critical temperature and Lifshitz point found previously by using the magnetic susceptibility \cite{B90, Z92}. The results discussed above show also the existence of the second $\chi'$  susceptibility maximum below $T_C$ for the field $B \leq 0.5$ mT (along the intermediate {\bf b} axis) which decreases, widens and shifts to higher temperature with increasing field. The existence of such a maximum which can be an indicator of a new enigmatic magnetic phase was reported  in pure MnP by Becerra et al. \cite{Bec1}, however, for the magnetic field one order of magnitude smaller. The feature that has not been reported in MnP is a sharp peak of $\chi''$ at the Lifshitz point. 

We have measured the field dependence of the isothermal magnetic coefficient for $B \parallel {\bf c}$ (easy axis) at several temperatures. It allows us to estimate the critical exponent $\omega$ which describes the critical singularity of $M_T \sim B^{-\omega}\mid_{ T=T_c}$ at zero field critical temperature. The obtained $\omega \approx 0.35$ is closer to the value from three dimensional Heisenberg model value $\omega \approx 0.364$ rather than to the expected Ising one ($\omega=0.433$). However, it should be noted that because of the technical reasons the measurements were performed at finite field and one can expect that for the field small enough the behavior of the system evolves from the Heisenberg (isotropic) into the Ising (anisotropic) one. 

To describe the high temperature part of the phase diagram - critical lines between disordered phase Para and ordered phases Ferro and Mod, we have proposed a simple phenomenological model. This is a hybrid in which the system of uniform layers is described by the anisotropic $s=1/2$ Heisenberg model, where the coefficients of the Landau free energy are found within the MFA. The appropriate order parameter is the component of the magnetization along the easy axis $m_x$. It has allowed us to find nine coefficients of the Landau free energy (6) as the functions of one internal parameter as well as reduced temperature and field. On the other hand, to take into account that the magnetization in several layers is different (Mod phase) we introduced an additional order parameter $q$ which is zero for the system of the layers with the same magnetization. The appropriate terms of the free energy have been introduced in the spirit of the Landau phenomenological theory, assuming that both order parameters $m_x$ and $q$ are small near LP. This leads to the additional coefficients $A_q$ and $B_q$ (15). According to the Landau idea we have assumed that $A_q$ is a linear function of the distance from, in this case, the Lifshitz point, and $A'$ (20) and $B_q$ are temperature and field independent near the LP. Such a procedure allows us to reconstruct the high temperature part of the phase diagram according to the experimental results with the critical temperature between Para and Ferro phase $t_c(h) \sim h^{2.1}$ (21), and between Para and Mod phase $t_c(h) \sim (h-h_{LP})^{0.6}$ (22). Our simple molecular-field theory leads also to qualitatively reasonable, as compared with experimental data, description of the temperature and field dependence of the magnetization, magnetic susceptibility and magnetocaloric effect. Particularly, the existence and shift with increasing field of the transverse susceptibility maximum in the paramagnetic phase and the step like Ferro/Para transition, which suggest that the inflection point in $M_T(B)$ curves can define the critical point. At the phase transition between Para and Mod phases $M_T$ changes sign which is an indicator of the existence of the antiferromagnetic correlations \cite{Sznajd}, and in this case the inflection point coincides with the sign change of $M_T(B)$.


\section{Acknowledgments}
Authors would like to thank H. Fjellv\aa g of University of Oslo for providing the Mn$_{0.9}$Co$_{0.1}$P single crystal used in this work. The research was supported by a Grant No. N N202 193234 of the Polish Ministry of Science and Higher Education.



\begin{thebibliography}{}
\bibitem{Reis} M.S. Reis, R.M. Rubinger, N.A. Sobolev, M.A. Valente, K. Yamada, K. Sato, Y. Todate, A. Bouravleuv, P.J. von Ranke, and S. Gama, Phys. Rev. B {\bf 77}, 104439 (2008).
\bibitem{Plac1} T. ~Plackowski, D. ~Kaczorowski, and Z. ~Bukowski, Phys. Rev. B {\bf 72}, 184418 (2005).
\bibitem{Plac2}  T. ~Plackowski, M. Matusiak, and J. ~Sznajd, Phys. Rev. B {\bf 82}, 094408 (2010).
\bibitem{Plac3} T. ~Plackowski, Y. X. ~Wang, and A. ~Junod, Rev. Sci. Instrum. {\bf 73}, 2755 (2002).
\bibitem{F84}  H.~Fjellvag, A. Kjekshus Acta Chem. Scand. A  {\bf 38}, 563 (1984);  H.~Fjellvag, A. Kjekshus, A.~Zieba, and S.~Foner,  J. Phys. Chem. Solid  {\bf 45}, 709 (1984);
\bibitem{F90} H.~Fjellvag, A. Kjekshus, and A.~Zieba, Acta Chem. Scand. A  {\bf 44}, 8 (1990).
\bibitem{B90} C.C.~Becerra, A.~Zieba, N.F.~Oliveira, Jr., and H.~Fjellvag, J. Appl. Phys. {\bf 67}, 9 (1990).
\bibitem{Z92} A.~Zieba, C.C.~Becerra, H.~Fjellvag, N.F.~Oliveira, Jr., and A. Kjekshus, Phys. Rev. B  {\bf 46}, 3380 (1992).
\bibitem{Bec1} C.C.~Becerra, Y. Shapira, N.F.~Oliveira, Jr., and T.S.~Chang, Phys. Rev. Lett.  {\bf 44}, 1692 (1980). 
\bibitem{Bec2} Y. Shapira, C.C.~Becerra,  N.F.~Oliveira, Jr., and T.S.~Chang, Phys. Rev. B {\bf 24}, 2780 (1981). 
\bibitem{Bec3} R.H.~Moon, J.M.~Cable, and Y. Shapira, J. Appl. Phys. {\bf 52}, 2025 (1989). 
\bibitem{Bec4} V.~Bindilatti, C.C.~Becerra,  and N.F.~Oliveira, Jr.,  Phys. Rev. B {\bf 40}, 9412 (1989). 
\bibitem{Bec5} C.C.~Becerra, N.F.~Oliveira Jr., and Y. Shapira, J. Physique Coll.  {\bf 49}, C8 895 (1988).
\bibitem{Yokoi} C.S.O.~Yokoi, M.D.~Coutinho-Filho, and S.R.~Salinas, Phys. Rev. B {\bf 29}, 6341 (1984).
\bibitem{Slota} Andrzej Zieba, Monika Slota, and Mariusz Kucharczyk, Phys. Rev. B {\bf 61}, 3435 (2000).
\bibitem{Huber} E.E. Huber Jr. and D.H. Ridgley, Phys. Rev. {\bf 135}, A1033 (1964).
\bibitem{Yos} H.~Yoshizawa, S.M.~Shapiro, and T.~Komatsubara, J. Phys. Soc. Jap.  {\bf 54}, 3084 (1985). 
\bibitem{Dob} L.~Dobrzynski and A.F.~Anderson, J. Magn. Magn. Mat.  {\bf 82}, 67 (1989).
\bibitem{Bec2000} C.C.~Becerra, J.Phys.: Condens. Matter {\bf 12}, 5889 (2000).
\bibitem{Yam} T.~Yamazaki, Y.~Tabata, T.~Waki, H.~Nakamura, M.~Matsuura, and N.~ Aso, J. Phys.:Conference Series {\bf 200}, 32079 (2010).
\bibitem{Hiyamizu} S.~Hiyamizu and T.~Nagamiya, Int. J. Magn. {\bf 2}, 33 (1972).
\bibitem{Bec6} C.C. Becerra, N.F. Oliveira Jr., and A.C. Migliano, J. Appl. Phys. {\bf 63}, 3092 (1988).
\bibitem{Sznajd} J. Sznajd, Phys. Rev. B {\bf 78}, 214411 (2008).
\bibitem{Plac4} T. ~Plackowski and D. ~Kaczorowski, Phys. Rev. B {\bf 72}, 224407 (2005).
\bibitem{Terui} H. Terui, T. Komatsubara and E. Hirahara, J. Phys. Soc. Jap.  {\bf 38}, 383 (1975).
\bibitem{Riedel} E. Riedel and F. Wegner, Z. Physik {\bf 225}, 195 (1969).
\bibitem{Pleimling} Michel Pleimling and Malte Henkel, Phys. Rev. Lett. {\bf 87}, 125702 (2001).
\bibitem{Diehl} H.W. Diehl, M. Shpot {\bf 62}, 12338 (2000).
\bibitem{Leite} Marcelo M. Leite {\bf 68}, 052408 (2003).
\bibitem{Sz} J. ~Sznajd, Phys. Rev. B {\bf 64}, 052401 (2001).
\bibitem{Chen} Y.~Xiang, Y.~Chen, Q.Z.~Chen, J.~Zhang, and Y.K. Liu, Solid State Commun. {\bf 148}, 369 (2008); Y.~Xiang, Y.~Chen, Q.Z.~Chen, J.~Zhang, and Y.K. Liu, J. Magn. Magn. Mat.  {\bf 321}, 163 (2009)




\end{thebibliography}
\end{document}